# Electric field control of magnetic states in isolated and dipole-coupled FeGa nanomagnets delineated on a PMN-PT substrate

Hasnain Ahmad[1], Jayasimha Atulasimha[2] and Supriyo Bandyopadhyay[3]

[1]Department of Electrical and Computer Engineering
[2]Department of Mechanical and Nuclear Engineering
Virginia Commonwealth University, Richmond, VA 23284, USA
Corresponding author email: sbandy@vcu.edu

ABSTRACT

We report observation of a "non-volatile" converse magneto-electric effect in elliptical FeGa nanomagnets delineated on a piezoelectric PMN-PT substrate. The nanomagnets are initially magnetized with a magnetic field directed along their nominal major axis and subsequent application of an electric field across the substrate generates strain in the substrate, which is partially transferred to the nanomagnets and rotates the magnetizations of some of them to metastable orientations. There they remain after the field is removed, resulting in "non-volatility". In isolated nanomagnets, the angular separation between the initial and final magnetization directions is $< 90^0$, but in a dipole-coupled pair consisting of one "hard" and one "soft" nanomagnet, the soft nanomagnet's magnetization is rotated by $> 90^0$ from the initial orientation because of the dipole influence of the hard nanomagnet. These effects can be utilized for ultra-energy-efficient logic and memory devices.

The ability to control the magnetization states of nanomagnets with an electric field is alluring fornanomagnetic logic and memory applications [1-8]. It also enables controlling nanoscale magnetic fields with an electric field, which is a technological advance in its own right [9, 10].

Electric field control of magnetization in thin film heterostructures of ferromagnetic/ferroelectric layers (artificial multiferroics) has been reported in the past by many groups [11-14]. This was later extended to microscale artifical multiferroics [15, 16] and then to nanoscale artificial multiferroics [17-19]. The latter [17-19] are of particular interest since, there, an electric field generates strain in a piezoelectric layer which is transferred to a (magnetostrictive) ferromagnetic layer in elastic contact with the piezoelectric layer, and this affects the ferromagnetic layer's magnetization. Obviously, this can have device applications in rewritable non-volatile memory if the ferromagnet persists in the new state and does not revert back to the old state after the electric field is withdrawn. Ref. [17] reported evolving the magnetization state of a nanoscale ferromagnet from a single domain state to a transitional domain state with an electric field. The ferromagnet, however, reverted to its original single domain state after the electric field was removed, making the effect "volatile'". This volatility was also a feature of ref. [18]. Here, we report a similar nanoscale converse magneto-electric effect in *FeGa* nanomagnets, but in our case, the state that the magnetization is driven to by the electric field is *metastable*. Consequently, when the electric field is withdrawn, the magnetization persists in the new state and does not spontaneously return to the original state, resulting in "non-volatility". Ref. [19] also reported a similar non-volatile transition, but there the final state of magnetization was stable instead of metastable. From the point of view of binary logic and memory applications, it makes no difference whether the final state is stable or metastable as long as the metastable state is robust against thermal perturbations (that is the case here since the nanomagnets driven to the metastable state persist in that state indefinitely at room temperature). However, driving the system from one stable state to a metastable state can be preferable to driving it to another stable state if the energy barrier separating the first stable state and the metastable state is lower than the energy barrier separating two stable states. In that case, the energy dissipated in the transition will be lower, making it more attractive for device applications.

There is one other feature that distinguishes our work from past work. All previous work on (electrically-generated) strain-induced nanomagnet switching [17-19] utilized elemental magnetostrictive nanomagnets made of Co or Ni. They have relatively small saturation magnetostriction and therefore would require relatively large electric fields to alter the magnetization state. The alloy FeGa, on the other hand, has much higher magnetostriction [20, 21] and is therefore more desirable, but it also presents unique material challenges owing to the existence of many phases [22, 23]. Nonetheless, there is a need to step beyond elemental ferromagnets and examine compound or alloyed ferromagnets with much higher magnetostriction to advance this field. There has been some work in FeGa thin films [24-27], but not in nanoscale FeGa magnets which are important for nanomagnetic logic and memory applications. This motivates our work.

Our system consists of elliptical FeGa nanomagnets of 200 - 350 nm feature size fabricated on a (100)-oriented PMN-PT substrate (70% PMN and 30% PT). The fabrication involved electron beam lithography and sputtering [see the accompanying supplementary material for details of fabrication and material characterization]. The major axes of the elliptical nanomagnets are aligned nominally parallel to each other on the substrate. Prior to delineation of the nanomagnets, the PMN-PT substrate is poled with an electric field of 9 kV/cm in a direction that will coincide with the major axes of the nanomagnets.

Fig. 1 shows atomic force (AFM) and magnetic force (MFM) micrographs of two nearly elliptical isolated nanomagnets that are placed far enough away from each other to make dipole interaction between them negligible. The nanomagnet nominal dimensions are: major axis = 335 nm, minor axis = 286 nm and thickness = 15 nm. Underneath each nanomagnet there is a ~5 nm layer of Ti needed to adhere the nanomagnets to the PMN-PT substrate. This layer is thin enough to allow most of the strain generated in the PMN-PT substrate (upon the application of an electric field) to transfer to the FeGa nanomagnets resting on top. There is no capping layer on the nanomagnets to prevent oxidation of FeGa since we found that to be a non-issue.

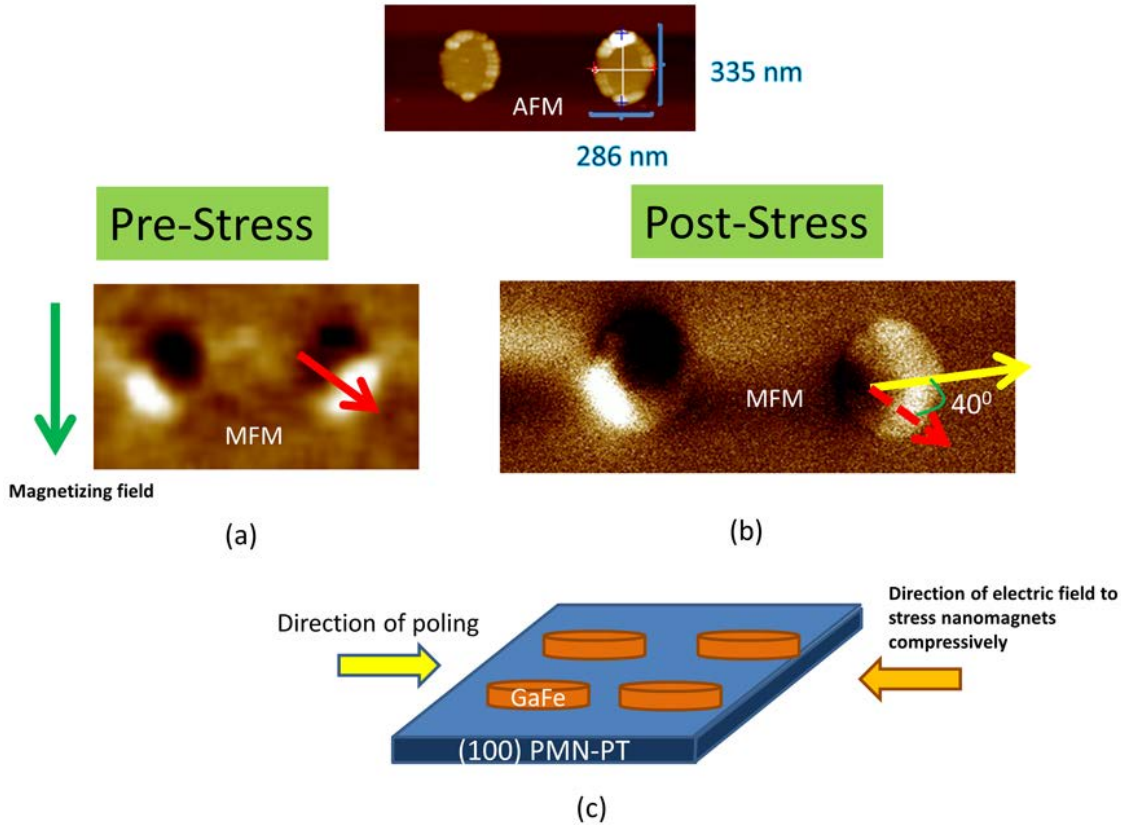

Figure 1: (a) Magnetic force (MFM) and atomic force (AFM) micrographs of isolated elliptical nanomagnets that have been subjected to a strong magnetic field in the direction indicated by the vertical green arrow. The resulting magnetization direction of the right nanomagnet is indicated by the slanted red arrow. (b) The MFM image of the nanomagnets after they have been stressed with an electric field and relaxed [stress withdrawn]. The magnetization direction of the left nanomagnet does not show any discernible change, but the right nanomagnet's magnetization has rotated to a new orientation shown by the slanted solid yellow arrow. For comparison, the initial orientation of this nanomagnet's magnetization is shown by the broken red arrow. The magnetization has rotated by ~$40^0$ owing to the stress generated by the electric field and subsequent removal of the electric field has not returned the magnetization to the original state, but left it in the new state. (c) The poling direction and direction of electric field applied to generate compressive stress along the nominal major axes of the nanomagnets.

The nanomagnets are initially magnetized in one direction along their major axes with a 1.5 Tesla magnetic field as shown by the vertical green arrow in Fig. 1(a). The MFM image in Fig. 1(a) shows that after the magnetizing field has been removed, the nanomagnets have close to a single-domain state with their magnetizations pointing not quite along that of the applied magnetic field, but reasonably close to it. The deviation could be due to lithography imperfections [see the AFM images for the magnet topography] that make the shapes of the nanomagnets slightly non-elliptical (which is why the stable orientation is not exactly along the major axis) or the presence of pinning that pins the magnetization in an orientation subtending a small angle with the major axis.

An electric field of 4.2 kV/cm is then applied across the PMN-PT substrate in the direction opposite to the poling direction to generate compressive stress along the nanomagnets' major axes (see Fig. 1(c)). This field is generated by applying a voltage of 2.1 kV across a 5 mm long substrate and this voltage is within the linear strain-versus-voltage regime determined in ref. [19]. This field strains the PMN-PT substrate owing to $d_{33}$ coupling. The value of $d_{33}$ measured in our substrates in ref. [19] was 1000 pm/V. Therefore, the average strain generated in the PMN-PT substrate is 420 ppm. If all of it is transferred to the FeGa nanomagnets, then the stress generated in them is ~42 MPa since the Young's modulus of FeGa is about 100 GPa.

The nanomagnets are compressed in the direction of their major axes by the electric field and since FeGa has a positive magnetostriction, this should rotate the magnetization of the magnetized nanomagnets toward the minor axis because stress anisotropy relocates the potential energy minimum to an orientation that is perpendicular to the stress axis. It is expected that after the electric field (or stress) is removed, the magnetization will return to the original orientation, or perhaps to some other orientation, since the potential energy landscape changes when stress is withdrawn.

What we see, however, in the MFM image of Fig. 1(b) [post-stress condition] is that the left nanomagnet's magnetization is indeed in the original orientation but the right nanomagnet's magnetization is not. The left's magnetization is in the original orientation because it either did not rotate at all [perhaps owing to the fact that the stress generated in it was insufficient to overcome the shape anisotropy energy barrier in this nanomagnet and make its magnetization rotate, or the magnetization was pinned by defects] or it did rotate but returned to the original orientation after the removal of stress, as expected. What is interesting is that the right nanomagnet's magnetization has assumed an orientation subtending an angle of ~40$^0$ with the original. Clearly, in this nanomagnet, stress was able to overcome the shape anisotropy energy barrier and budge the magnetization from its original orientation, but after stress withdrawal, the magnetization settled into a different orientation and stayed there. This is obviously a metastable orientation corresponding to a local potential minimum that is robust against thermal noise and could have arisen either because the shape of the right nanomaget deviates significantly from elliptical causing local minima to appear in the potential profile of the nanomagnet, or pinning sites pin the magnetization in a given state after stress withdrawal, or there are multiple phases that lead to multiple coercivities [28] associated with the presence of multiple energy barriers in the potential profile separated by local energy minima. The stress anisotropy gives rise to an effective magnetic field given by $H_{eff} = (3/2)|\lambda_s|\sigma/(\mu_0 M_s)$, where $\lambda_s$ is the magnetostriction coefficient, $\sigma$ is the stress, $\mu_0$ is the permeability of free space and $M_s$ is the saturation magnetization of the nanomagnet. If the material has two different coercivities $H_1$ and $H_2$ and the inequality $H_1 < H_{eff} < H_2$ holds, then stress can get the magnetization stuck in a metastable state. In the accompanying supplementary material, we show the magnetization curves of the nanomagnets and they do seem to indicate the presence of multiple phases with different coercivities. Therefore, this scenario is very likely.

There can be another feature peculiar to FeGa alloys. A recent paper suggests that there is significant non-Joulian magnetostriction in FeGa alloys [29] which can complicate the stress-induced magnetization rotation process in this materialand perhaps drive the magnetization to metastable states.

No matter what the underlying mechanism is, in this case, the magnetization state of the right nanomagnet has been altered with an electric field resulting in evolution of the magnetization from one orientation to another. Equally important, this transformation is "non-volatile" because the magnetization does not return to the original state after removal of the electric field.

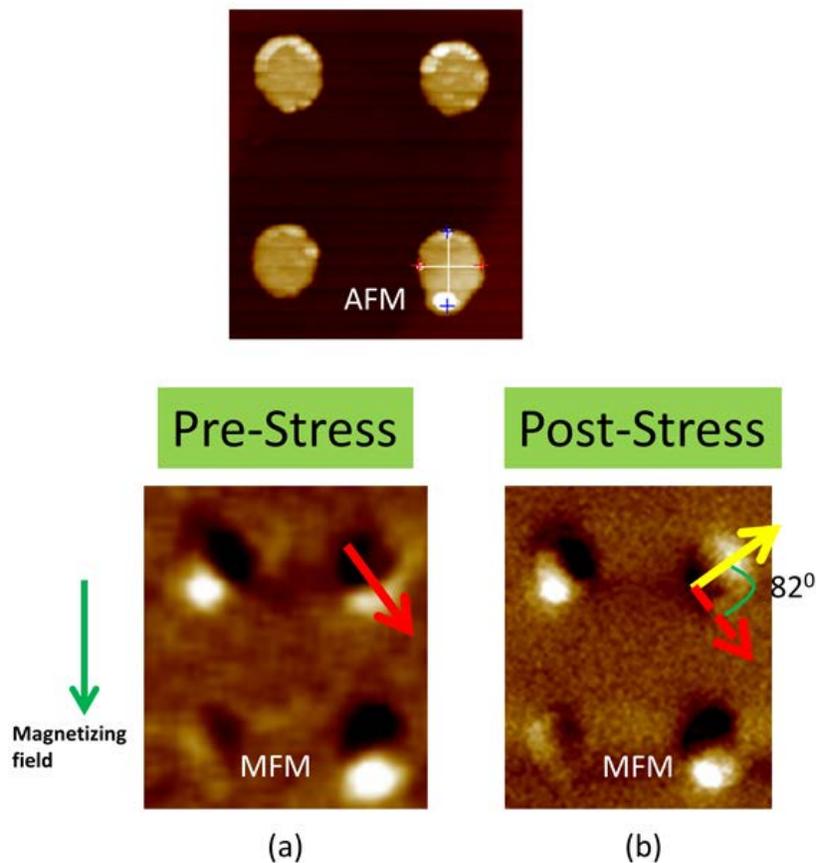

Figure 2: (a) Same as Fig. 1(a), except there are four nanomagnets in this image. The initial magnetization direction of the nanomagnet in the upper right corner is indicated by the slanted red arrow. (b) MFM image after application and removal of stress. The magnetizations of all nanomagnets expect the one in the upper right corner show no discernible difference between the pre- and post-stress conditions, but the magnetization of the one in the upper right corner has rotated by ~$82^0$. Notice that this nanomagnet is most "circular" of all and therefore has the lowest shape anisotropy energy barrier, which is why stress was able to rotate its magnetization.

To ascertain that this effect is repeatable, we examined another set of nanomagnets of slightly different dimensions [major axis = 337 nm, minor axis = 280 nm and thickness = 16 nm]. The corresponding AFM and MFM pictures are shown in Fig. 2. The three nanomagnets (among the four shown) that have the highest shape anisotropy (highest eccentricity of the ellipse) do not show any perceptible difference between the pre- and post-stress magnetization states [either because their magnetizations did not rotate when stressed or returned to the original states after stress removal], but the least shape anisotropic nanomagnet has evolved to a different state after stress removal. The new state's magnetization has an angular separation of ~$82^0$ from the original magnetization state. Once again, the new state is non-volatile. Altering the magnetization state of a magnet with an electric field is the converse magnetoelectric effect. Therefore, we have observed clear evidence of the converse magnetoelectric effect in the nanoscale and the effect has the property of *non-volatility*.

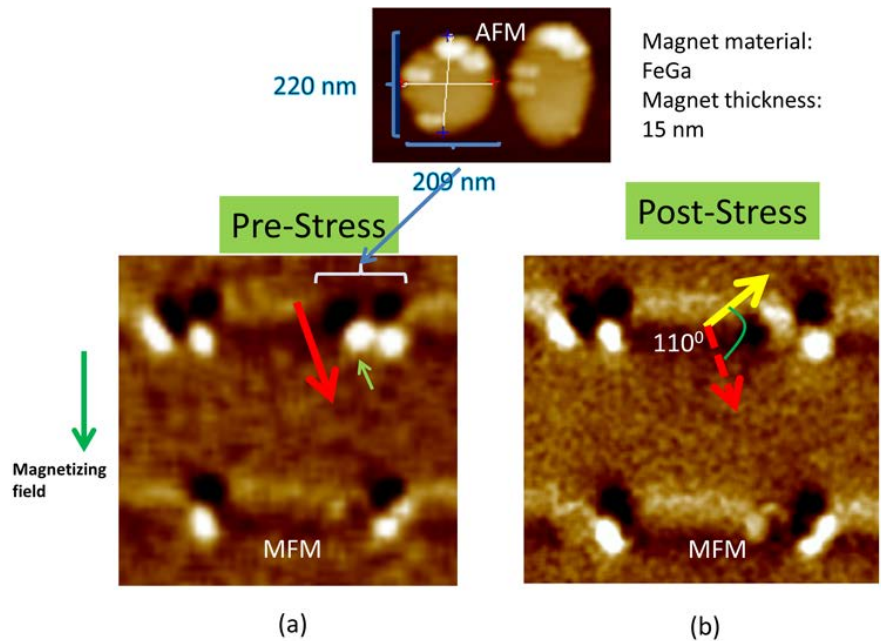

Figure 3: Dipole-coupled pairs where the right partner is much more shape anisotropic than the left partner, making the right partner hard and the left partner soft. (a) MFM image after being magnetized by a magnetic field in the direction of the green vertical arrow and before application of stress. (b) MFM image after application and removal of stress. The magnetizations of all pairs, expect the one in the upper right corner show no discernible difference between the pre- and post-stress conditions, but the magnetization of the soft nanomagnet in the upper right corner (indicated by the short light green arrow) has rotated by ~$110^0$.

In order to investigate whether this effect is influenced by dipole coupling between nanomagnets, we fabricated arrays of closely spaced nanomagnet pairs whose mutual separations are small enough to allow reasonable dipole coupling between them. One nanomagnet in the pair is intentionally made much more shape anisotropic than the other. This makes the former magnetically stiff (or hard) and its magnetization cannot be budged by the stress generated because of the extremely high shape anisotropy energy barrier. The latter has much less shape anisotropy and hence is softer. Its magnetization state can be affected by stress.

Fig. 3(a) shows the pre-stress MFM images of four such pairs that have all been initially magnetized with a strong magnetic field to make their magnetizations approximately parallel to each other. Dipole coupling would prefer anti-parallel ordering within a pair, but the dipole coupling strength is insufficient to overcome the shape anisotropy energy barrier of any soft magnet to rotate its magnetization and make the two magnetization orientations in a pair mutually anti-parallel after the magnetizing field has been removed. Next, all nanomagnets are compressed along their nominal major axes with an electric field and the field is withdrawn. Fig. 3(b) shows the post-stress MFM images. Three pairs show no perceptible difference between the initial and final orientations, but the fourth pair in the upper right hand corner shows that the magnetization of the soft nanomagnet has rotated by a larger angle of ~$110^0$ (> $90^0$) and once again the new state is non-volatile. The larger angle of rotation (compared to the case of isolated nanomagnets) is obviously due to dipole coupling. The latter prefers anti-ferromagnetic ordering within a pair, i.e., the magnetization of the soft nanomagnet should be anti-parallel to that of the hard nanomagnet. This is not completely achieved because the magnetization ultimately gets trapped in a metastable state, but it does not get trapped in a metastable state that is *close* to the original state because the dipole coupling is strong enough to dislodge the magnetization from any such state and steer it to a state subtending a large angle with the original state. The difference between the isolated and dipole-coupled cases is that in the former, the angular separation between the old and new states is less than $90^0$, whereas in the latter, it is greater than $90^0$. Therefore, the converse magnetoelectric effect is influenced by dipole coupling.

In Fig. 4, we show two more dipole coupled pairs, initially magnetized in the same direction, where the magnetization of the soft nanomagnet has rotated by ~$150^0$ in one case, and ~$180^0$ in the other case, after application and withdrawal of stress. Once again the dipole coupling, which prefers anti-parallel magnetizations of the two magnets, is responsible for the >$90^0$ rotation of the magnetization.

In conclusion, we have demonstrated a non-volatile converse magneto-electric effect in the nanoscale which is affected by dipole coupling. These results not only show that electric field control of local magnetization orientation (and hence local magnetic field) is possible, but also that the effect is non-volatile and can be influenced by neighboring magnetization states because of dipole coupling. This influence is critical to implement the conditional dynamics of Boolean (or even non-Boolean) logic where the state of one logic device determines the state of the next. The dipole-coupled pair, for example, acts as an inverter (NOT gate) which is clocked by the electric field (the application of the field makes the magnetizations of the partners nearly anti-parallel, so that if binary bits are encoded in magnetization orientation, then the output bit becomes the logic complement of the input). Finally, because the magnetization rotation is non-volatile, there are also potential applications in non-volatile memory where bits, encoded in the magnetization orientation, are written with a voltage.

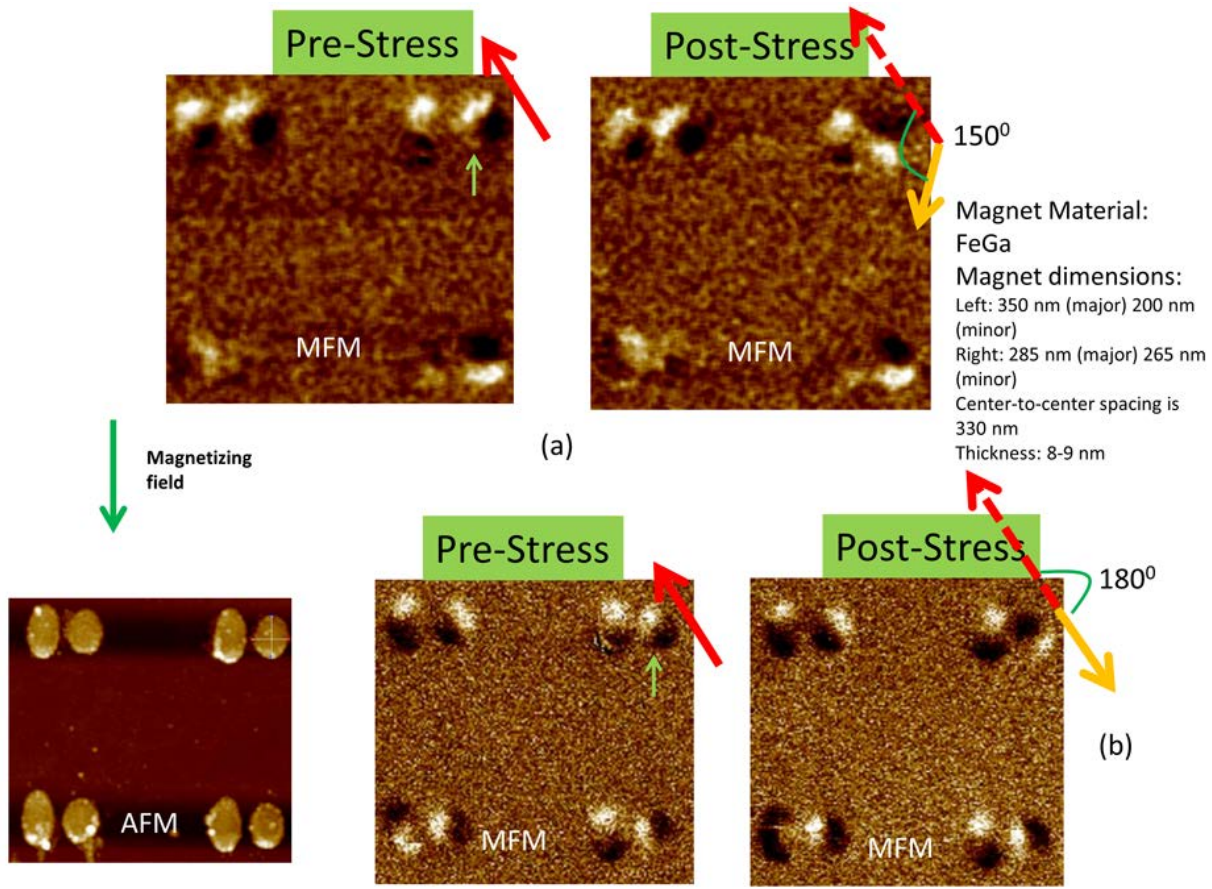

Figure 4: (a) MFM image of a dipole-coupled pair after being magnetized by a magnetic field in the direction of the green vertical arrow and before application of stress is shown on the left. In these pairs, the left partner is more shape-anisotropic than the right, i.e., the left nanomagnet is "hard" and the right nanomagnet is "soft". MFM image after application and removal of stress is shown on the right. The magnetizations of all expect the pair in the upper right corner show no discernible difference between the pre- and post-stress conditions, but the magnetization of the soft nanomagnet in the upper right corner (indicated by the short light green arrow) has rotated by ~$150^0$. (b) Similar to (a), except now the rotation is by $180^0$.

There are challenges however, because clearly the effect is not observed in every fabricated specimen, but in a minority of specimens. The possible causes for the poor yield are: (i) lithography imperfections that make the shape anisotropy energy barrier and the potential profile different in different nanomagnets, (ii) random material defects resulting in pinning sites that trap the magnetization and prevent its rotation under stress in some nanomagnets, and (iii) multiple phases in FeGa that spawn different energy barriers and different potential profiles in different nanomagnets. All this points to a serious challenge; very precise lithography will be required to translate these observations into a viable technology and material defects need to be mitigated.

Nonetheless, because of the extremely low energy dissipation in electric field control of magnetization [3,4,30,31], these initial observations are encouraging for future energy-efficient nanomagnetic computing.

## ACKNOWLEDGMENTS

This work was supported by the US National Science Foundation under grants ECCS-1124714 and CCF-1216614. The sputtering of FeGa nanomagnets was carried out at the National Institute of Standards and Technology, Gaithersburg, Maryland, USA.

# Supplementary Material

**Fabrication of nanomagnets on poled PMN-PT substrates**

The PMN-PT substrates were first poled with an electric field of 8000 V/cm in a direction which will coincide with the major axes of the nanomagnets (the major axes were nominally parallel to each other). The poled substrates were spin-coated with two layers of PMMA (e-beam resist) of different molecular weights in order to obtain superior nanomagnet feature definition: PMMA-495 Anisol and PMMA-950 Anisol at 2500 rpm spinning rate. The resists were then baked at $115^0$ C for 2 minutesand exposed in a Hitachi SU-70 SEM with a Nabity attachment using 30 kV accelerating voltage and 60 pA beam current. Subsequently, the resists were developed in MIBK:IPA (1:3) for 70 seconds followed by rinsing in cold IPA.

For nanomagnet delineation, a 4-5 nm thick Ti layer was first deposited using e-beam evaporation at a base pressure of $(2-3) \times 10^{-7}$ Torr, followed by the deposition of 12-17 nm of FeGa (thickness verified with AFM) using DC magnetron sputtering of a FeGa target with a base pressure of $(2-3) \times 10^{-8}$ Torr and deposition pressure of 1 mTorr. The magnetron power was 45 W and the deposition was carried out for 45 seconds. The nanomagnets were formed following lift-off and they were imaged with both scanning electron microscope (SEM) and atomic force microscope (AFM).

Magnetization states were ascertained with magnetic force microscopy (MFM). All MFM imaging was carried out with a low moment MFM tip in order to perturb the magnetization states of the nanomagnets as little as possible.

**X-ray diffraction**

Grazing incidence and in-plane x-ray diffraction (XRD) were carried out to ascertain the crystallographic orientation of the grains and compared to data available in the literature. Figure

S1 shows the XRD data for the nanomagnets. The film is polycrystalline containing [110], [200] and [211] oriented grains. The in-plane data indicate <100> textured growth [<110> texturenormal to the film plane] as also reported in ref. [S1].

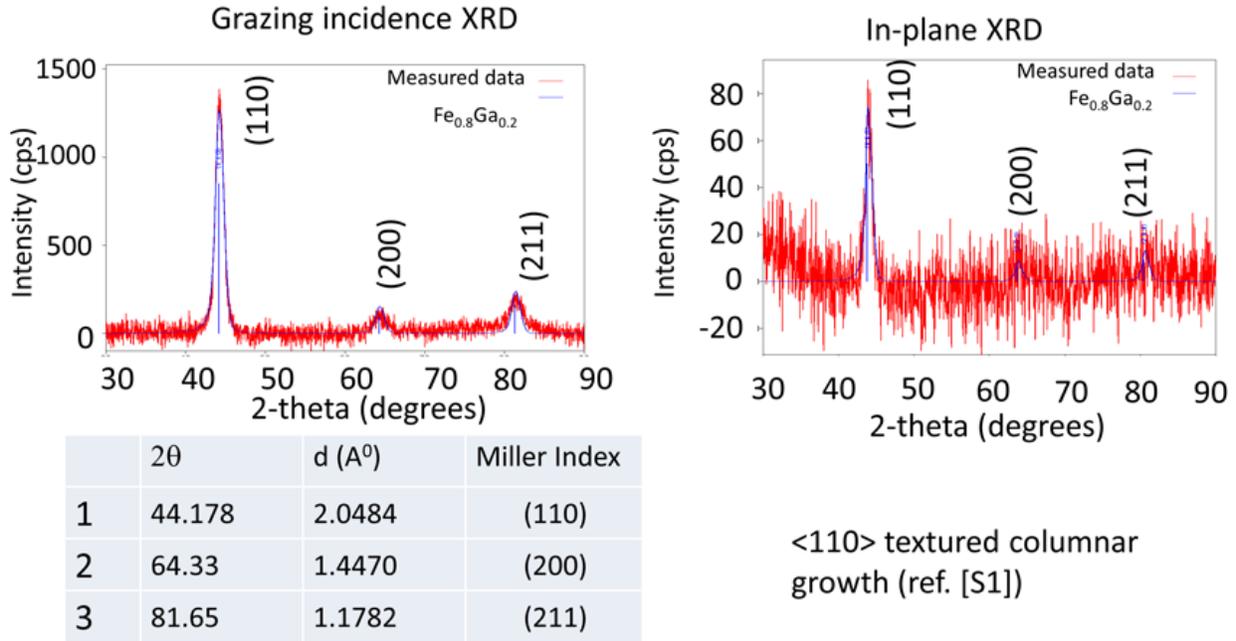

Figure S1: The grazing angle and in-plane x-ray diffraction data showing that the sputtered nanomagnets are polycrystalline with [110], [200] and [211] oriented grains. The in-plane data indicate <110> textured growth.

**X-ray photoelectron spectroscopy**

A well-known problem with sputter deposited FeGa films is the variation of Fe and Ga atomic percent from one layer to another [S1-S3]. To study this compositional variation in our nanomagnets, we carried out x-ray photoelectron spectroscopy (XPS) by successively removing layers with Ar ion etching to obtain signal from various layers to ascertain the elemental composition of each layer. The etching is stopped once we obtain a signal from the substrate (the substrate was [100]-silicon in this case), indicating that we are approachingthe bottom layer.

Since FeGa oxidizes in air, we get a signal from oxygen. In Figure S2(a) we show the elemental composition as a function of etch time or layer depth obtained from XPS. In Figure S2(b) we show the relative abundance of Fe and Ga in each layer after suppressing the oxygen signal. As expected, the oxygen content is highest at the topmost layer and dwindles as we approach the bottom (interface with the substrate). The Fe mole fraction also increases with depth. Because of the compositional variation, we expect the effective magnetostriction to be less than the value reported for $Fe_{0.8}Ga_{0.2}$ in the literature.

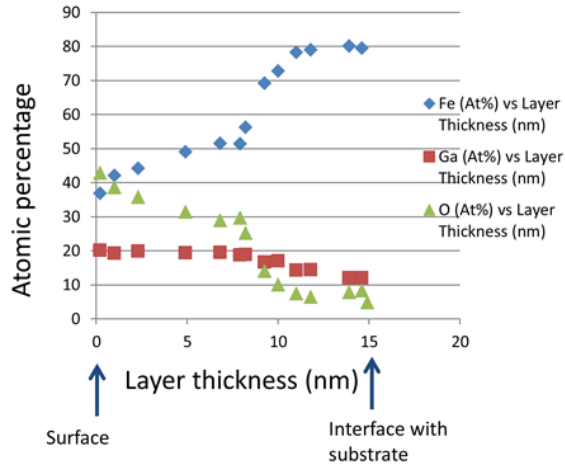 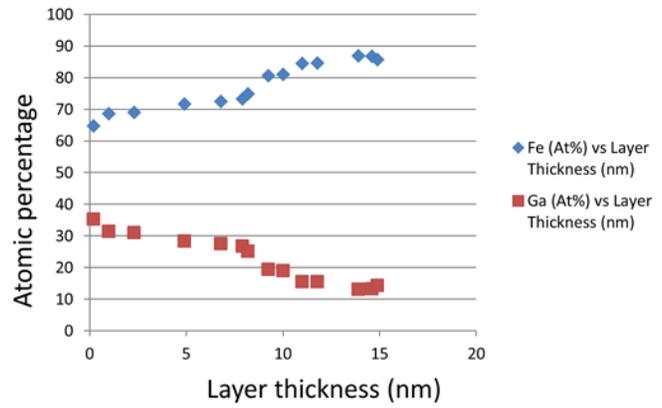

Figure S2: XPS data showing layer by layer composition.

**Film morphology**

In Figure S3, we show a scanning electron micrograph of the film surface. This is very similar to that shown in ref. [S1]

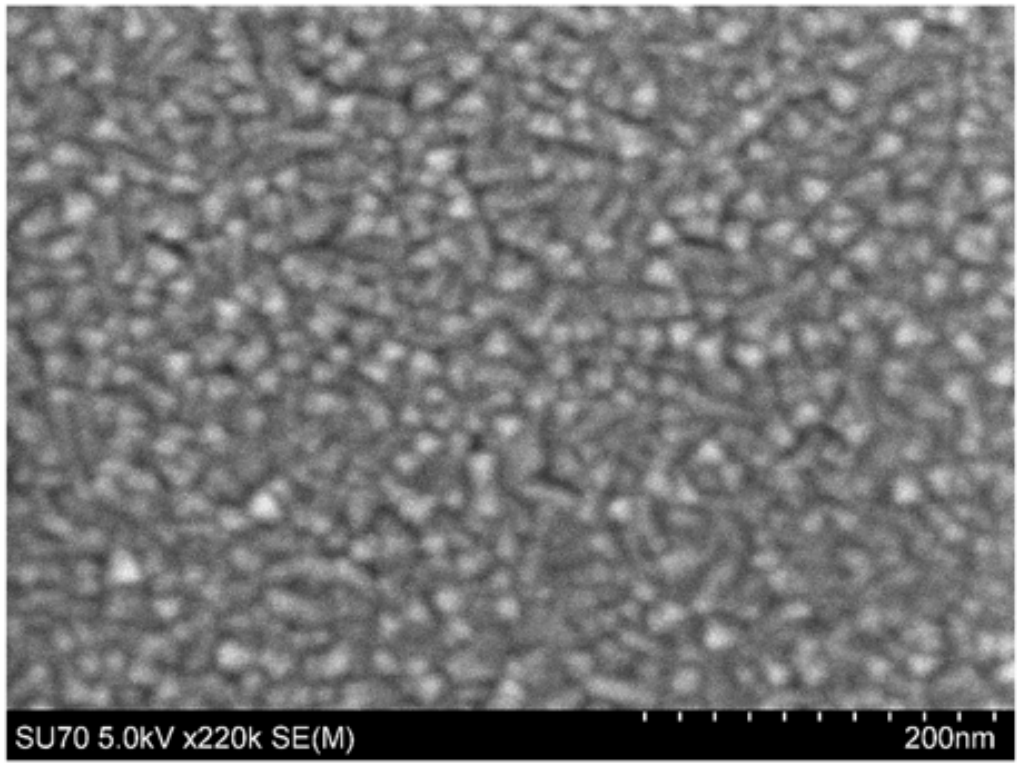

Figure S3: Scanning electron micrograph of the film surface showing the morphology.

**Magnetization studies**

The magnetization (M-H) curves were obtained at 77 K and 300 K in a Quantum Design Vibrating Sample Magnetometer with the magnetic field in-plane and perpendicular-to-plane. These results are shown in Figure S4. In Figure S5 we zoom in on the low field characteristic at 300 K to determine the coercivity of the film and to examine if there are multiple coercivities associated with multiple phases. The shape of the curve (there is a broad shoulder) indicates that there are at least two coercivities associated with different phases. Similar behavior was observed in ref. [S1].

The presence of multiple coercivities indicates the presence of multiple energy barriers which will spawn local energy minima and hence metastable states. This is very likely the reason why metastable states exist and the magnetization gets stuck in one of these states when driven by stress out of a stable state. In ref. [19], which dealt with an elemental ferrromagnet Co instead of a compound ferromagnet like FeGa, metastable states were not found and the Co nanomagnets exhibited a single coercivity. Therefore, the metastable states appear to be a material feature. From the standpoint of device applications, whether the magnetization is driven by stress to a stable state or a metastable state is immaterial, as long as the initial and final states are magnetically distinct.

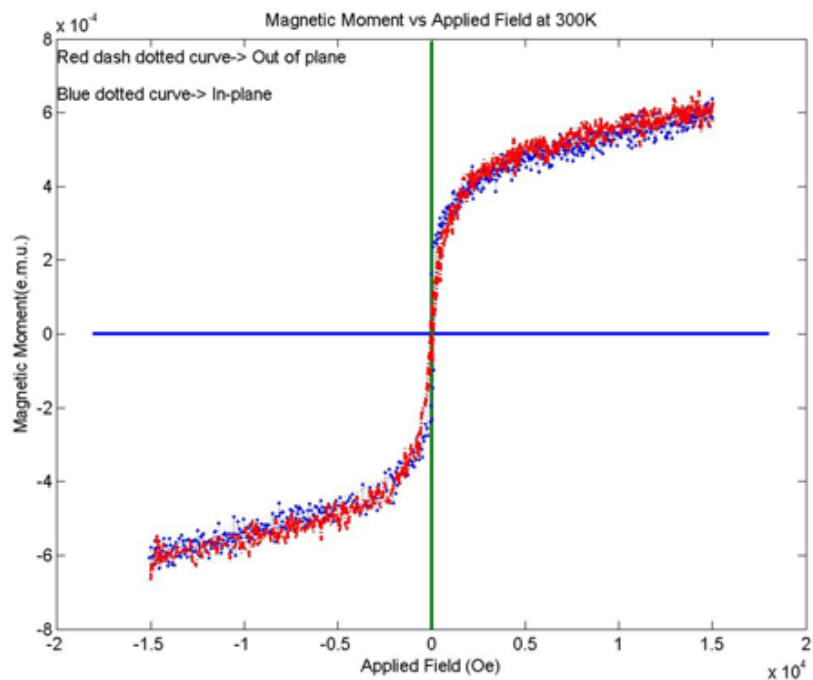

(a)

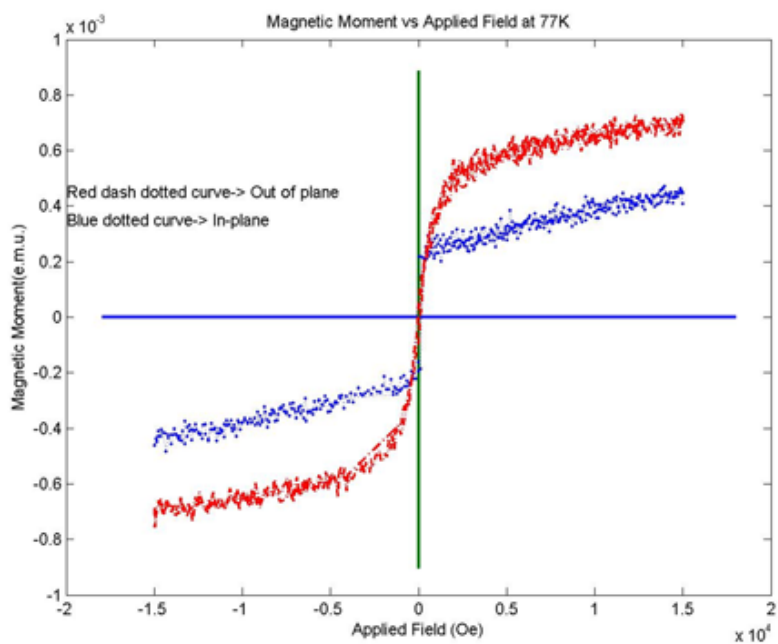

(b)

Figure S4: Magnetization curves with the magnetic field in-plane and perpendicular-to-plane. The results are plotted for two different temperatures.

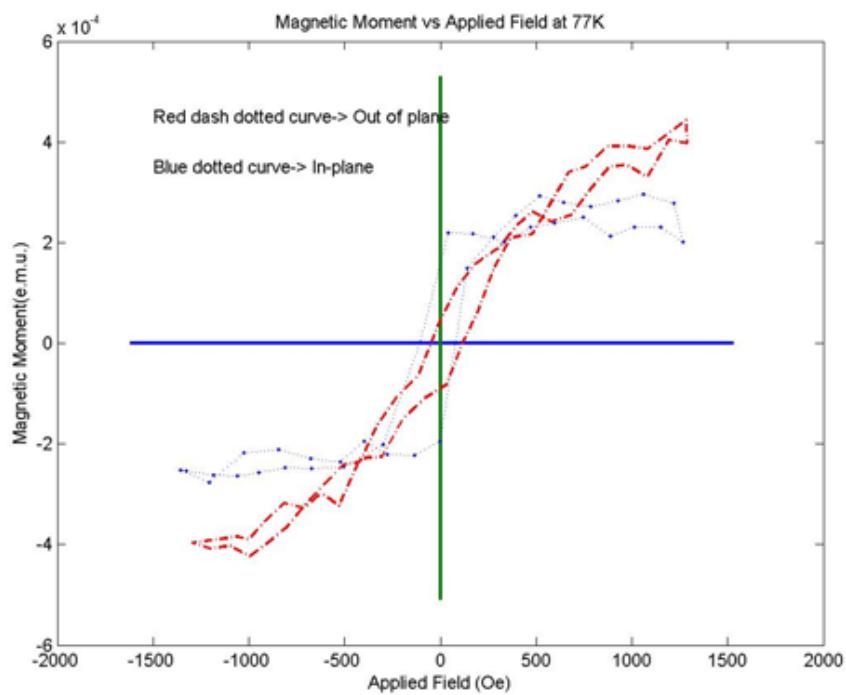

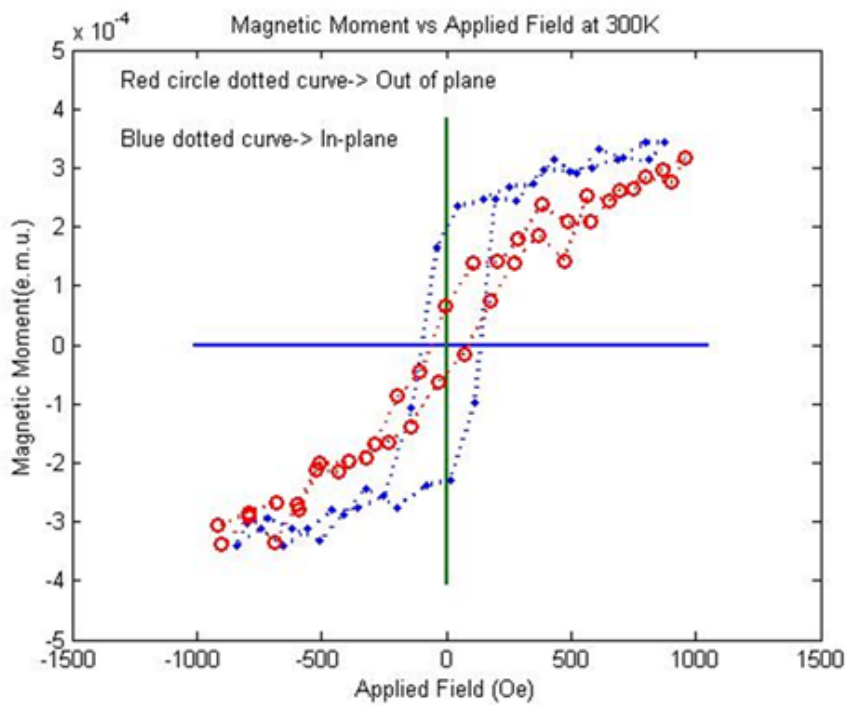

Figure S5: Magnetization curve at low magnetic fields at 77 K and 300 K. The shape of the curve bears telltale sign of multiple phases with multiple coercivities as observed before in ref. [S1].